Modeling the Impact of Interventions on an Epidemic of Ebola in Sierra Leone and Liberia

Caitlin M. Rivers[1], Eric T. Lofgren[1], Madhav Marathe[1], Stephen Eubank[1], Bryan L. Lewis[1]

[1]Network Dynamics and Simulation Science Lab, Virginia Bioinformatics Institute, Virginia Polytechnic Institute and State University, Blacksburg, VA USA

Corresponding Author:
Caitlin M. Rivers
Network Dynamics and Simulation Science Laboratory
Virginia Bioinformatics Institute
Virginia Polytechnic Institute and State University
1880 Pratt Drive (0477)
Blacksburg, VA 24061 USA
Tel: 540-724-1596
Email: cmrivers@vbi.vt.edu

**Abstract**

*Background*
An Ebola outbreak of unparalleled size is currently affecting several countries in West Africa, and international efforts to control the outbreak are underway. However, the efficacy of these interventions, and their likely impact on an Ebola epidemic of this size, is unknown. Forecasting and simulation of these interventions may inform public health efforts.

*Methods*
We use existing data from Liberia and Sierra Leone to parameterize a mathematical model of Ebola and use this model to forecast the progression of the epidemic, as well as the efficacy of several interventions, including increased contact tracing, improved infection control practices, the use of a hypothetical pharmaceutical intervention to improve survival in hospitalized patients.

*Findings*
Model forecasts until Dec. 31, 2014 show an increasingly severe epidemic with no sign of having reached a peak. Modeling results suggest that increased contact tracing, improved infection control, or a combination of the two can have a substantial impact on the number of Ebola cases, but these interventions are not sufficient to halt the progress of the epidemic. The hypothetical pharmaceutical intervention, while impacting mortality, had a smaller effect on the forecasted trajectory of the epidemic.

*Interpretation*
Near-term, practical interventions to address the ongoing Ebola epidemic may have a beneficial impact on public health, but they will not result in the immediate halting, or even obvious slowing of the epidemic. A long-term commitment of resources and support will be necessary to address the outbreak.

*Funding*: This work was funded by NIH MIDAS Grant 5U01GM070694-11 and DTRA Grant HDTRA1-11-1-0016 and DTRA CNIMS Contract HDTRA1-11-D-0016-0001.

## Introduction

Western Africa is currently experiencing an unprecedented outbreak of Ebola, a viral hemorrhagic fever. On March 23, 2014 the World Health Organization announced through the Global Alert and Response Network that an outbreak of Ebola virus disease in Guinea was unfolding[1-3]. Ebola is generally characterized by sporadic, primarily rural outbreaks, and has not been seen before in West Africa, or in an outbreak of this size.

As of September 14, 2014, the World Health Organization has reported 5,335 cases of Ebola virus disease in Sierra Leone, Liberia, Guinea and Lagos, Nigeria[4]. Considerable attention has been focused on preventing the outbreak from spreading further, either within Africa or intercontinentally. In principle many of the measures to contain the spread of Ebola, such as intensive tracing of anyone in contact with an infected individual and the use of personal protective equipment (PPE) for healthcare personnel treating infected cases, are straightforward[5]. However, implementing those interventions in a resource poor setting in the midst of an ongoing epidemic is far from simple, and subject to a great deal of uncertainty.

Mathematical models of disease outbreaks can be helpful under these conditions by providing forecasts for the development of the epidemic that account for the complex and non-linear dynamics of infectious diseases and by projecting the likely impact of proposed interventions before they are implemented. This in turn provides policy makers, the media, healthcare personnel and the public health community with timely, quantifiable guidance and support[6-9].

We use a mathematical model to describe the development of the Ebola outbreak to date, provide projections for its future development, and examine the potential impact of several interventions, namely increased contact tracing, improved access to PPE for healthcare personnel, and the use of a pharmaceutical intervention to improve survival in hospitalized patients.

## Methods

*Outbreak Data*

A time series of reported Ebola cases was collected from public data released by the World Health Organization, as well as the Ministries of Health of the afflicted countries. These data sets do not include patient level information and reflect reported, rather than laboratory confirmed cases, but they represent the best available data on the outbreak. A curated version of this data is available at https://github.com/cmrivers/ebola.

*Compartmental Model*

A compartmental model was used to describe the natural history and epidemiology of Ebola, adapted from Legrand *et al.*[10] which was previously used to describe the 1995 Democratic Republic of Congo and 2000 Uganda Ebola outbreaks. Briefly, the population is divided into six compartments, as shown in Figure 1. Susceptible individuals (S) may become Exposed (E) after contact with an infectious individual and transition in turn to the Infectious (I) class after the disease's incubation period, thereafter capable of infecting others. A proportion of these individuals may be Hospitalized (H). Both untreated patients in I and hospitalized patients in H may experience one of two outcomes: patients may die, with a chance of infecting others during the resulting funeral (F) before being removed from the model (R), or they may recover, at which point they are similarly removed. The system of ordinary differential equations describing this model is available in the Web Material posted to Figshare.

**Figure 1. Compartmental flow of a mathematical model of the Ebola Epidemic in Liberia and Sierra Leone, 2014.** The population is divided into six compartments: Susceptible (S), Exposed (E), Infectious (I), Hospitalized (F), Funeral (F) – indicating transmission from handling a diseased patient's body, and Recovered/Removed (R). Arrows indicate the possible transitions, and the parameters that govern them. Note that λ is a composite of all β transmission terms described in Table 1.

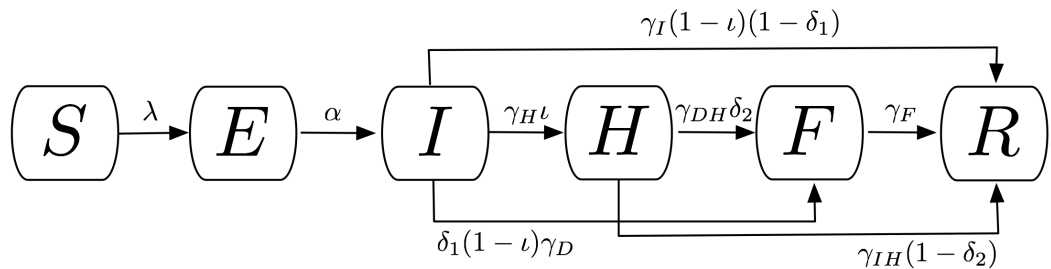

*Model Fitting and Validation*

A deterministic version of the model was fit and validated against the current outbreak data using weighted least-squares optimization, with seed values from the Uganda outbreak described in Legrand *et al*[10]. The last 15 days of reported cases were given one-quarter of the weight in the model, to capture improved case detection and the bulk of cases. Based on anecdotal reports from the field (unpublished), candidate optimized fits were accepted such that roughly one-quarter of infections each came from contacts with hospitalized patients or funereal transmission, with the balance being from person-to-person spread within the community. The optimizer was further constrained to plausible parameter values, such as an upper bound of 20 days for infection duration and time from death to burial, and 0 to 1 for probabilities or proportions. This model was fit only for Sierra Leone and Liberia, as the outbreak in Guinea has a unique epidemic curve, which necessitates an alternative model design beyond the scope of this paper. The fitted parameters for this model, as well as their descriptions, may be found in Table 1.

**Table 1. Model Parameters and Fitted Values for a Model of an Ebola Epidemic in Liberia and Sierra Leone, 2014.**

| Parameter | Liberia Fitted Values | Sierra Leone Fitted Values |
|---|---|---|
| Contact Rate, Community ($\beta_I$) | 0.160 | 0.128 |
| Contact Rate, Hospital ($\beta_H$) | 0.062 | 0.080 |
| Contact Rate, Funeral ($\beta_F$) | 0.489 | 0.111 |
| Incubation Period ($1/\alpha$) | 12 days | 10 days |
| Time until Hospitalization ($1/\gamma_H$) | 3.24 days | 4.12 days |
| Time from Hospitalization to Death ($1/\gamma_{DH}$) | 10.07 days | 6.26 days |
| Duration of Traditional Funeral ($1/\gamma_F$) | 2.01 days | 4.50 days |
| Duration of Infection ($1/\gamma_I$) | 15.00 days | 20.00 days |
| Time from Infection to Death ($1/\gamma_D$) | 13.31 days | 10.38 days |
| Time from Hospitalization to Recovery ($1/\gamma_{IH}$) | 15.88 days | 15.88 days |
| Daily probability a Case is Hospitalized ($\iota$) | 0.197 | 0.197 |
| Case Fatality Rate, Unhospitalized ($\delta_1$) | 0.500 | 0.750 |
| Case Fatality Rate, Hospitalized ($\delta_2$) | 0.500 | 0.750 |

This validated model provides a mathematical description of the epidemic up to the present. In order to forecast into the future, a stochastic version of the model was implemented using Gillespie's algorithm with a tau-leaping approximation[11,12], which treats individuals as discrete units and converts the deterministic rates in the calibration model into probabilities, allowing random chance to come into play. Using the parameters from the calibration model, as well as the number of individuals in each compartment at the present date, 250 simulations of this model were run until December 31, 2014, giving a fan of potential epidemic trajectories that accounts for uncertainty in the forecast due to chance[13]. All models were implemented in Python 2.7, and the stochastic simulations used the StochPy library[14].

*Modeled Interventions*

Based on interventions that are technically, but not necessarily socially, feasible in the foreseeable future, we model five scenarios to examine their likely impact on the development of the epidemic. First, we model improved contact tracing by increasing the proportion of infected cases that are diagnosed and hospitalized from the baseline scenario of 51% in Liberia and 58% in Sierra Leone to 80%, 90% and

100%, and a concordant decrease in the time it takes for an infected individual to be hospitalized by 25%. Second, we explore the impact of simultaneously (1) decreasing the contact rate for hospitalized cases ($\beta_H$) to represent the increased use of PPE as supplies and awareness of the outbreak increase as well as (2) eliminating the possibility of post-mortem infection from hospitalized patients due to inappropriate funereal practices. Third, we model both a simultaneous (1) decrease in $\beta_H$ (lack of post-mortem infection from hospitalized cases) and (2) increase in the proportion of hospitalized cases. This models the effect of a joint, intensified campaign to identify and isolate patients – the conventional means of containing an Ebola outbreak – with the necessary supplies and infrastructure to treat these patients using appropriate infection control practices. Finally, we model a pharmaceutical intervention that increases the survival rate of hospitalized patients by 25%, 50% and 75%, with a moderately high level of contact tracing (80%).

*Human Subjects*

As this study uses purely publically available data without personal identifiers, it was determined not to require IRB approval.

**Results**

*Model Fit and Prediction*

The deterministic model fit well for both Liberia and Sierra Leone, with the predicted curve of cumulative cases following the reported number of cases in both countries. The end-of-year forecast shows a range of uncertainty for each country of several thousand cases between the most optimistic and pessimistic scenarios. However, in both, the number of cumulative cases is forecast to continue rising extremely rapidly, with the bulk of the epidemic yet to come (Figure 2). This suggests an extremely poor outlook for the course of the epidemic without intensive intervention on the part of the international community.

In the baseline end-of-year forecasts for both Sierra Leone and Liberia, person-to-person transmission within the community made up the bulk of transmission events, with a median (IQR: Interquartile Range) of 117877 (115100–120585) cases arising from the community in the Liberia forecast and 30611 (29667 – 31857) in the forecast for Sierra Leone. Both had fewer hospital transmissions – 21533 (21025 – 21534) in Liberia and 5474 (5306 – 5710) in Sierra Leone, than transmissions arising from funerals – 35993 (35163 – 36789) in Liberia and 9768 (9470 – 10137) in Sierra Leone. For brevity, only the results of the Liberia model are reported below, with the results from Sierra Leone in the electronic supplement. The epidemic trajectories for all modeled interventions may also be found in the Web Material.

**Figure 2. Fitted Compartmental Model for Ebola Epidemic in Liberia and Sierra Leone, 2014, with 250 Iterations of a Stochastic Forecast to December 31, 2014.** Red dots depict the reported number of cumulative cases of Ebola in each country, with the black line indicating the deterministic model fit. Each blue line indicates one of two hundred and fifty stochastic simulated forecasts of the epidemic, with areas of denser color indicating larger numbers of forecasts.

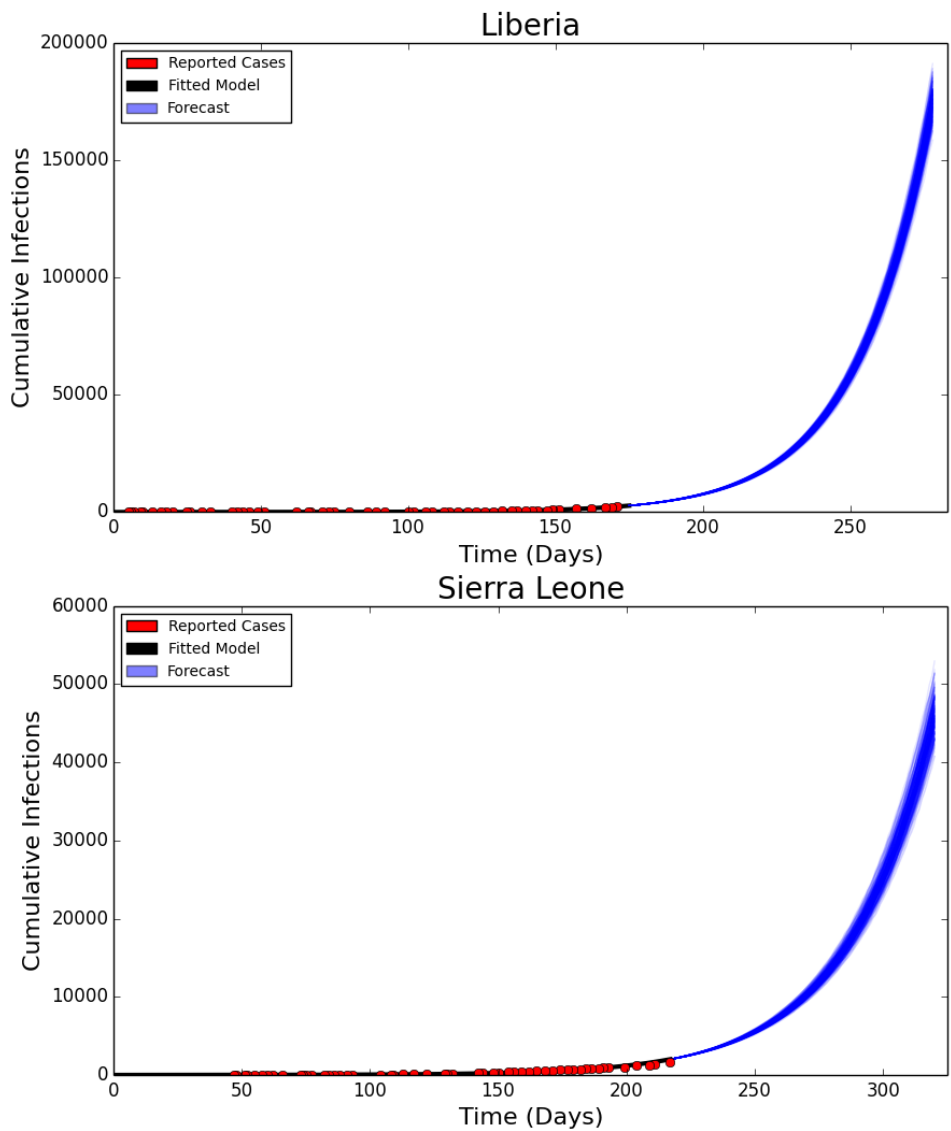

*Intensified Contact Tracing and Infection Control*

The forecast distribution of cases under intensified contact tracing is shown in Figure 3. As might be expected, there is a shift from community transmission toward hospital transmission, though at extremely high levels of contact tracing and hospitalization, the impact of the intervention on the course of the outbreak also results in fewer hospitalized cases. There is a less pronounced but still substantial downward shift in funeral cases, and a decrease in total cases in Liberia. This decrease is not however, sufficient to shift the cumulative case curve off its steep upward trajectory, but rather only lessens its magnitude.

The improved infection control scenario decreased $\beta_H$ to represent decreased risk of hospital transmission due to increased PPE, increased number of healthcare workers or greater awareness of the epidemic resulting in greater care while treating patients with undiagnosed febrile illness. Additionally, it eliminated the potential for post-mortem transmission during the funereal process. This combination of interventions resulted in a marked decrease in the overall number of cases (Figure 4).

Major reductions in all types of cases were seen, with the most dramatic drop in the relative number of cases arising from the reduction of within-hospital transmissions. However, even with substantially reduced transmission and a decrease in the burden of mortality from the outbreak, improved infection control was also insufficient to push the epidemic off its steep upward trajectory.

Figure 5 shows the median decrease in cases, as compared to the median of the baseline, no intervention scenario, for simulations combining increased contact tracing and a reduction in the risk of hospital transmission from those who are isolated and treated. The most optimistic of these scenarios, with complete contact tracing and a 75% reduction in hospital transmission results in more than 165,000 fewer total cases over the course of the forecasted period, as compared to the baseline scenario (Figure 6). This represents a ten-fold reduction in the number of cases, and is a major improvement in public health. However, even under this scenario, the epidemic is slowed and mitigated, rather than fully stopped, with new cases still arising after the end of the year.

**Figure 3. Distribution of Forecast Cases of Community, Hospital, Funeral and Total Cases for Ebola Epidemic, Liberia, 2014, at Baseline, 80%, 90% and 100% of Patients Traced and Hospitalized.** Box plots depict the median, interquartile range and 1.5 times the interquartile range for each scenario. Each individual simulated forecast is shown as a single dot, jittered so as to depict the complete distribution of the data.

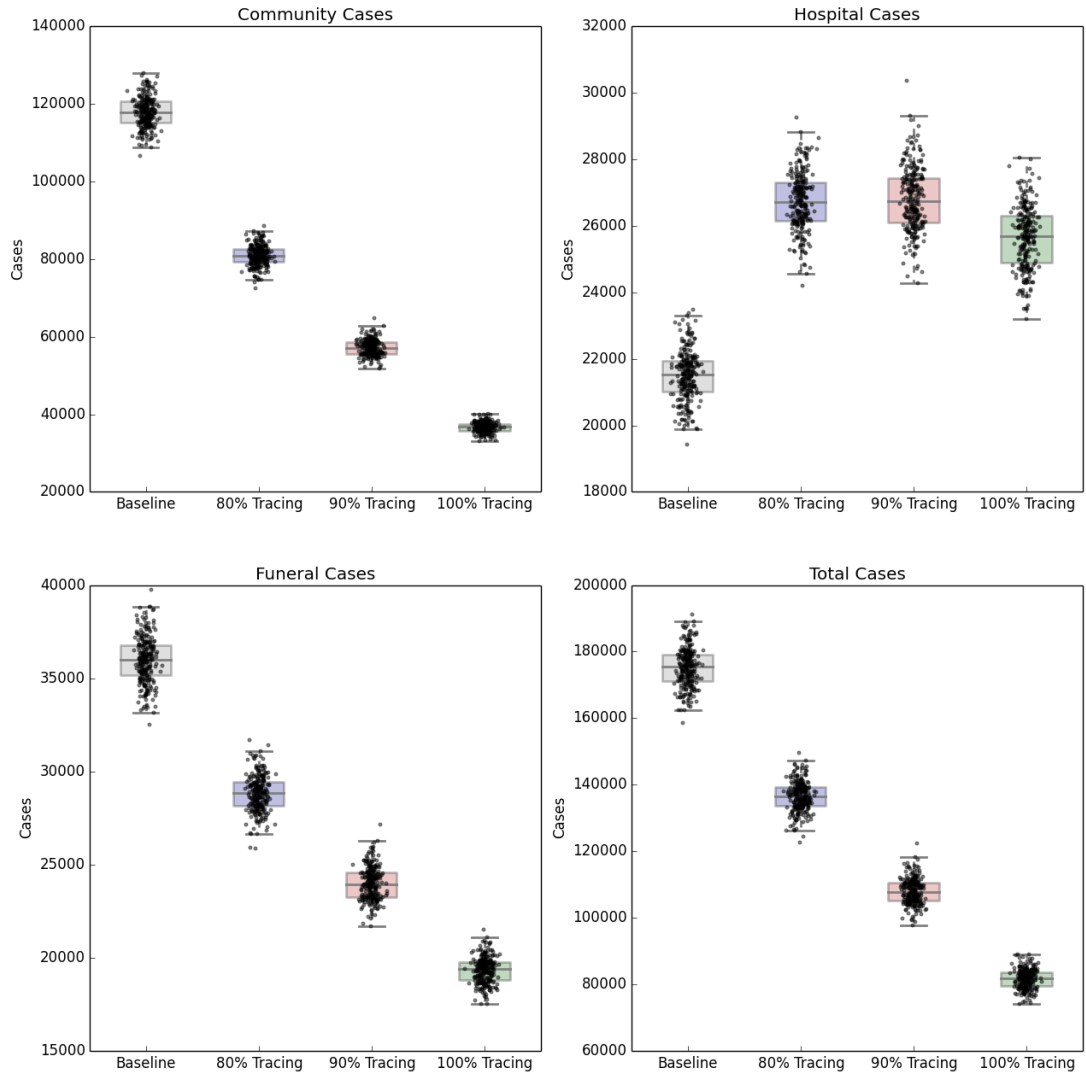

**Figure 4. Distribution of Forecast Cases of Community, Hospital, Funeral and Total Cases for Ebola Epidemic, Liberia, 2014, at Baseline, 25%, 50% and 75% Reductions in Hospital Transmission Contact Rates ($\beta_H$).** Box plots depict the median, interquartile range and 1.5 times the interquartile range for each scenario. Each individual simulated forecast is shown as a single dot, jittered so as to depict the complete distribution of the data.

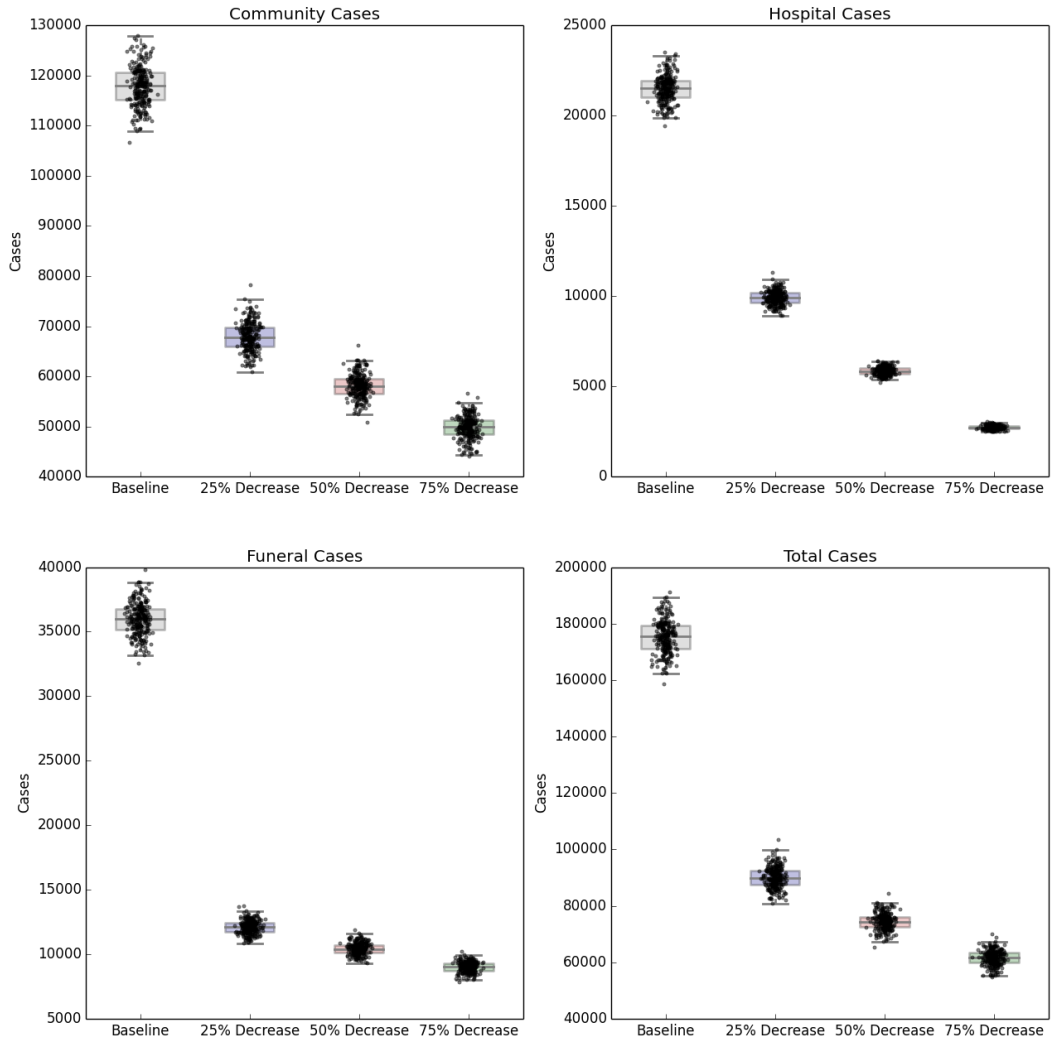

**Figure 5. Distribution of Forecasted Cases of Community, Hospital, Funeral and Total Cases for Ebola Epidemic, Liberia, 2014, at Baseline, 25%, 50% and 75% Reductions in Hospital Transmission Contact Rates ($\beta_H$) with 80%, 90% and 100% of Patients Traced and Hospitalized.** Each box represents the median result of 250 forecasted epidemics, each with a % of contacts traced and a % decrease in hospital transmission. Areas of deeper blue indicate progressively greater reductions of the median number of cases.

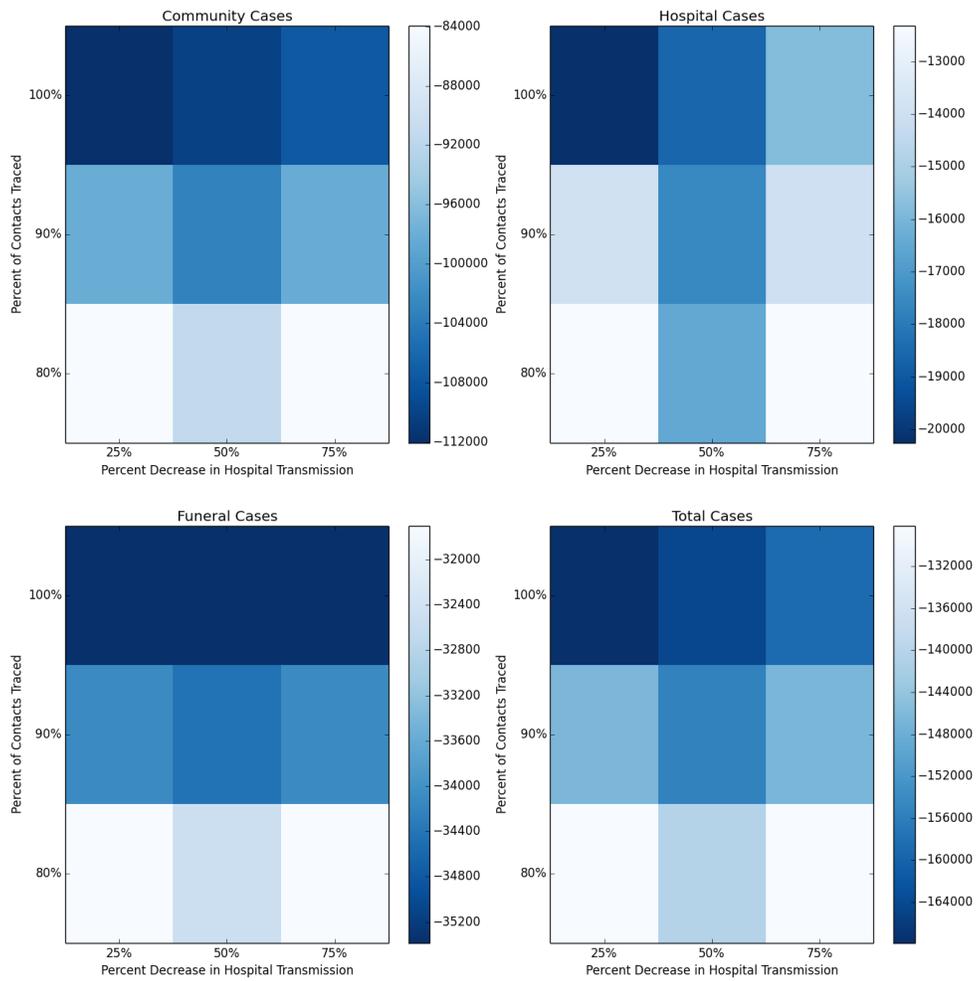

**Figure 6. Forecasted Cumulative Cases for Ebola Epidemic, Liberia, 2014 with 75% Reduction in Hospital Transmission Contact Rates ($\beta_H$) with 100% of Patients Traced and Hospitalized.** The solid black line represents the deterministic model fit of the epidemic to present, with each grey line representing a single simulated forecast with no interventions in place, and each blue line representing a single simulated forecast of the epidemic with 100% of contacts traced, a 75% (reduction in hospital transmission ($\beta_H$) and no post-mortem infections from hospitalized patients. Areas of darker color indicate more forecasts with that result.

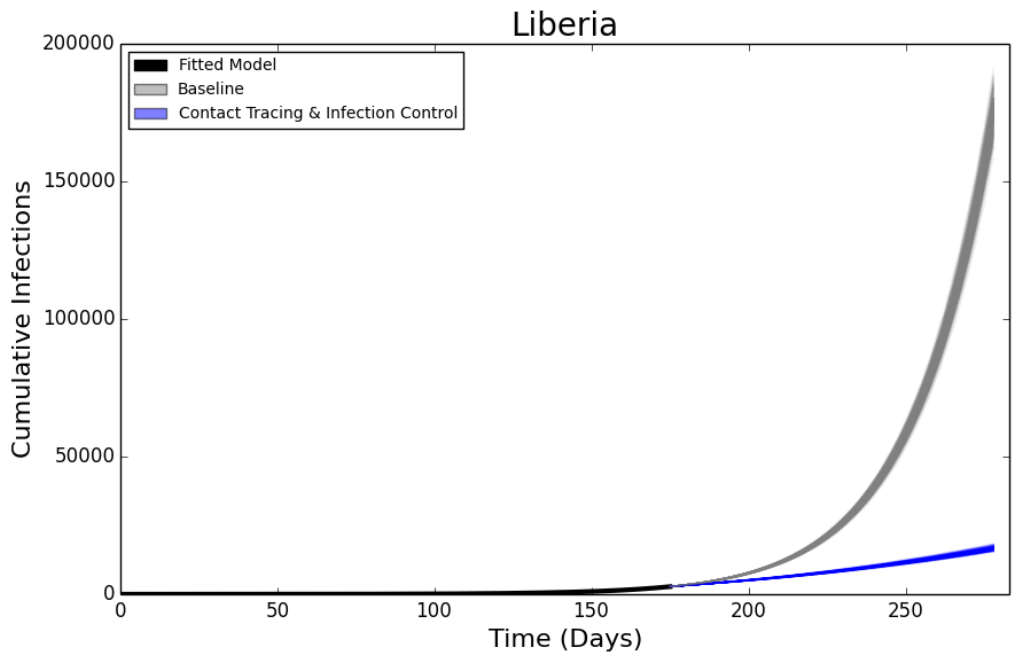

*Increased Availability of Pharmaceutical Interventions*

The widespread introduction of a pharmaceutical intervention that dramatically improves the survival rate of hospitalized patients also leads to a less severe outbreak, shown in Figure 7. Compared to contact tracing alone, there is a small reduction in the number of hospitalized cases (as the scenario implies no change in infection control practices), but a stronger decrease in the number of community, funeral and overall cases depending on the efficacy of the hypothetical pharmaceutical. As with the other forecasts above, this intervention also fails to halt the progress of the epidemic, though it does considerably reduce the burden of disease.

**Figure 7. Distribution of Forecast Cases of Community, Hospital, Funeral and Total Cases for Ebola Epidemic, Liberia, 2014, at Baseline, 25%, 50% and 75% Reductions in Case Fatality Rate Due to a Hypothetical Pharmaceutical Intervention.** Box plots depict the median, interquartile range and 1.5 times the interquartile range for each scenario. Each individual simulated forecast is shown as a single dot, jittered so as to depict the complete distribution of the data.

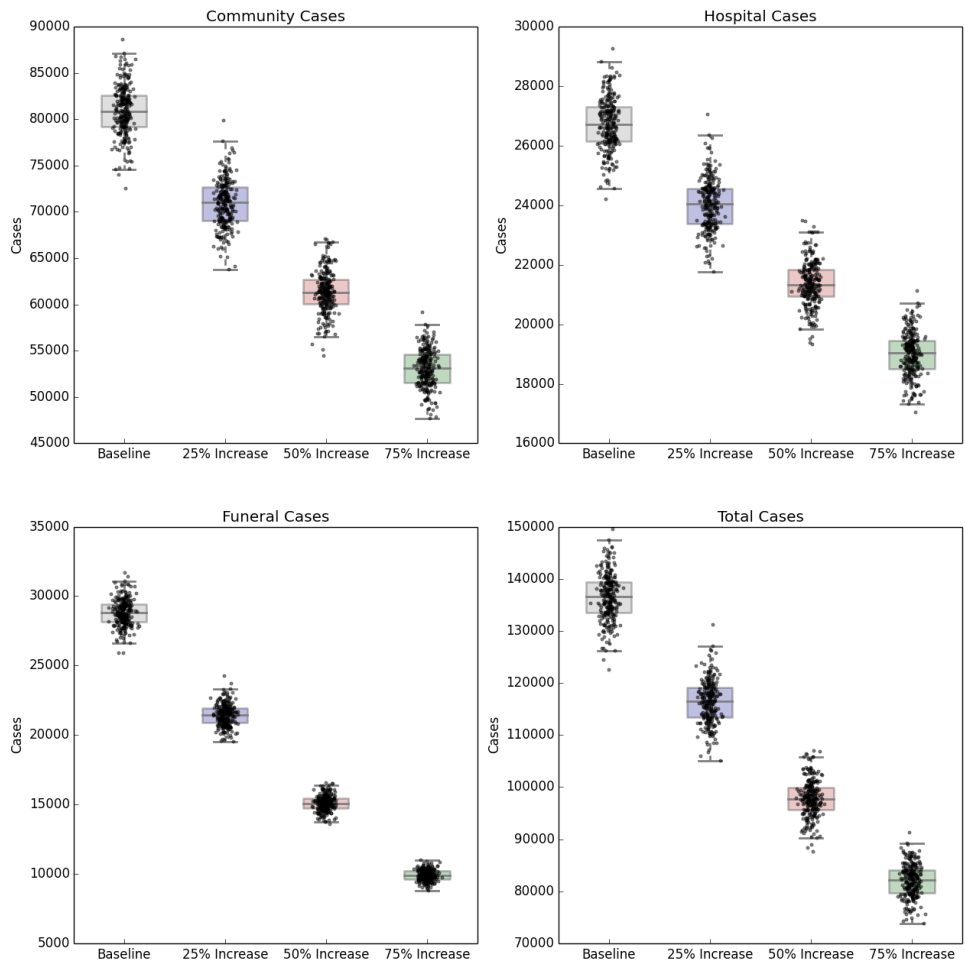

## Discussion

The control of Ebola outbreaks in the past has been a straightforward, albeit difficult application of infection control and quarantine policies. In principle, these types of interventions should be applicable to this outbreak as well, however it remains unclear whether or not they can be implemented at the unprecedented scale of the current outbreak. This study attempts to address whether or not interventions implemented at such a scale could arrest, or at least mitigate, the epidemic.

Our findings suggest that, at least in the near term, some form of coordinated intervention is imperative. The forecasts for both Liberia and Sierra Leone in the absence of any major effort to contain the epidemic paint a bleak picture of its future progress, suggesting we are presently in the opening phase of the epidemic, rather than near its peak.

Of the modeled interventions applied to the epidemic, the most effective by far is a combined strategy of intensifying contact tracing to remove infected individuals from the general population and place them in a setting that can provide both isolation and dedicated care, along with ensuring that such clinics have the necessary supplies, training and personnel to follow the necessary infection control practices. While both of these interventions in isolation also have an impact on the epidemic, they are much more effective in parallel, finding infected patients and seeing that they get the care they need to address their illness, and to prevent transmission to others either within or outside the hospital. In particular, the slight increase in cases in Sierra Leone at the lower end of the modeled contact tracing range, when unaccompanied by a concordant increase in infection control, highlights the necessity of these two interventions being implemented side-by-side.

The hypothetical mass application of a novel pharmaceutical, such as the one administered to two American aid workers [15] had a much smaller impact on the course of the epidemic itself. While certainly lessening the burden of mortality of those infected (in the most optimistic scenario modeled, reducing the case-fatality rate from 50% to 12·5%), the downstream effects of such an intervention are relatively minor, as there is no suggestion that any candidate treatments have a substantial impact on transmission. This will impact not only the evaluation of those treatments, which should thus focus primarily on their patient-level efficacy, but also in communicating to the media and the public that, despite the use of these drugs, the benefits that will be seen from them are in decreases to mortality due to infection, not in a halt to the epidemic itself.

Despite the considerable impact the proposed interventions have on the burden of disease, none of them are forecast, at least in the short term, to halt the epidemic entirely. It is possible these interventions will have a longer-term impact on the epidemic, however we have avoided projecting out further than the end of the year due to the inherent uncertainty in an emerging epidemic. This in turn suggests another communication challenge for public health planners. While all of the proposed interventions are worth pursuing, and will have an impact on the epidemic and public health, the attention of the international community must be sustained in the long term in order to ensure the necessary supplies and expertise remain present in the affected areas. Additionally, in light of public resistance to the limited types of these interventions already in place[16], public awareness and acceptance of intensified interventions must be built, as there will not be an immediate cessation to the epidemic, or even necessarily a clear sign that the situation is improving.

This study is not without limitations. As with all mathematical models, the results of the study depend on the assumptions about the natural history of Ebola, its epidemiology, and the values of the parameters used – as well as the quality of the data used to fit the model. Particularly, this model is validated against data that records cases on their time of reporting, rather than their time of onset, so the model time series may be shifted by several days. Because of the relatively small number of historical Ebola outbreaks, the unusual size of this outbreak, and the difficulty collecting data during an emerging epidemic, the accuracy of this model and its parameters is difficult to ascertain. It does however represent our best understanding of the epidemic using available data, and the model has proven capable of predicting the ongoing development of the epidemic, as well as having been used to model previous Ebola outbreaks[10]. The uncertainty inherent in model prediction has been addressed with the use of stochastic simulation to aid in quantifying uncertainty inherent within the model system.

The ongoing Ebola epidemic in West Africa demands international action, and the results of this study support that many of the interventions currently being implemented or considered will have a positive impact on reducing the burden of the epidemic. However, these results also suggest that the epidemic has progressed beyond the point wherein it will be readily and swiftly addressed by conventional public health strategies. The halting of this outbreak will require patient, ongoing efforts in the affected areas and the swift control of any further outbreaks in neighboring countries.

**Declarations of Interests**
The authors declare they have no financial or personal relationships with other people or organizations that may bias this work.

**Author Contributions**
All authors contributed to the study concept and design. CMR did the initial data collection and curation. BLL and CMR performed model fitting and evaluation. ETL was responsible for stochastic simulation and drafted the report. All authors reviewed, revised and approved the final version.

**Acknowledgements**
The authors would like to thank Katie Dunphy, Jesse Jeter, P. Alexander Telionis, James Schlitt, Jessie Gunter and Meredith Wilson for their assistance and support. The authors would also like to acknowledge the participants in the weekly briefings organized by DTRA, BARDA and NIH, including Dave Myer, Aiguo Wu, Mike Phillips, Ron Merris, Jerry Glashow, Dylan George, Irene Eckstrand, Kathy Alexander and Deena Disraelly for their comments and feedback.